\documentclass[12pt,preprint]{aastex}

\newcommand{\gtsim}{\protect\raisebox{-0.5ex}{$\:\stackrel{\textstyle >}
        {\sim}\:$}}

\begin{document}

\title{ An HI Threshold for Star Cluster Formation in
Tidal Debris\footnotemark[1]
\footnotetext[1]{Based in part on observations with the NASA/ESA
Hubble Space Telescope, obtained at the Space Telescope Science
Institute, which is operated by the Association of Universities for
Research in Astronomy, Inc., under NASA contract NAS5-26555}}

\author{Aparna Maybhate\altaffilmark{2}, Joseph Masiero\altaffilmark{3},
John E. Hibbard\altaffilmark{4}, Jane C. Charlton\altaffilmark{5},}
\author{Christopher Palma\altaffilmark{5}, Karen A. Knierman\altaffilmark{6},
and Jayanne English\altaffilmark{7}}

\altaffiltext{2}{Space Telescope Science Institute, 3700 San Martin Drive,
Baltimore, MD 21218; maybhate@stsci.edu}
\altaffiltext{3}{Institute for Astronomy, 2680 Woodlawn Dr,
Honolulu, HI 96822; masiero@ifa.hawaii.edu}
\altaffiltext{4}{National Radio Astronomy Observatory, 520 Edgemont
Road, Charlottesville, VA 22903; jhibbard@nrao.edu}
\altaffiltext{5}{Department of Astronomy and Astrophysics,
525 Davey Laboratory, Pennsylvania State University, University
Park, PA 16802; charlton@astro.psu.edu,
cpalma@astro.psu.edu}
\altaffiltext{6}{Steward Observatory, University of Arizona,
933 North Cherry Avenue, Tucson, AZ 85721; kknierman@as.arizona.edu}
\altaffiltext{7}{Department of Physics and Astronomy, University of
Manitoba, Winnipeg, Manitoba, Canada R3T 2N2;
jayanne$\_$english@umanitoba.ca}

\begin{abstract}
Super star clusters are young, compact star clusters found in the
central regions of interacting galaxies. Recently, they have also been
reported to preferentially form in certain tidal tails, but not in
others.  In this paper, we have used 21 cm HI maps and the {\it Hubble
Space Telescope} Wide Field Planetary Camera 2 images of eight tidal
tail regions of four merging galaxy pairs to compare the kiloparsec
scale HI distribution with the location of super star clusters found
from the optical images.  For most of the tails, we find that
there is an increase in super star cluster density with increasing projected HI
column density, such that the star cluster density is highest when log
$N_{HI} \gtsim\ 20.6$ cm$^{-2}$, but equal to the background count
rate at lower HI column density. However, for two tails (NGC 4038/39 Pos A and NGC 3921),
there is no significant star cluster population despite the presence
of gas at high column density.  This implies that the $N_{HI}$
threshold is a necessary but not sufficient condition for cluster
formation. Gas volume density is likely to provide a more direct
criterion for cluster formation, and other factors such as gas
pressure or strength of encounter may also have an influence.
Comparison of HI thresholds needed for formation of
different types of stellar structures await higher resolution HI
and optical observations of larger numbers of interacting galaxies.

\end{abstract}



\section{Introduction}

Massive star clusters tend to form in regions of vigorous star
formation, especially in starbursts triggered by galaxy interactions
and mergers \citep{sch98}. Images obtained with the {\it Hubble Space
Telescope} of several mergers and merger remnants show a large number
of young star clusters in the central regions
\citep{hol92,whi93,whi99,whi95,sch96,mil97,zep99}.  While detailed
studies of the central regions of merging galaxies have been
conducted, it is only recently that young star clusters have been
found in the tidal debris of some of these mergers
\citep{kni03,tra03,deg03,bas05}.  Using $V$- and $I$-band {\it HST}/WFPC2
images of the tidal tails of four merging pairs of galaxies,
\citet{kni03} find a statistically significant increase in the surface
density of pc-size star clusters on the optical tails 
compared to off-tail regions. 
The colour-magnitude diagrams (CMDs) of these sources differed from the
off-tail population, and are concentrated in the region of the CMD where
young star clusters are expected. 
In this small sample, those tails with a modest or no overdensity of
star clusters each contain an embedded candidate stellar tidal dwarf galaxy (TDG),
while the tails in the system with the highest star cluster density
(NGC 3256) lack such a structure.
This led \citet{kni03} to speculate that star
formation in tidal tails may manifest itself either in numerous small
structures like clusters along the tail or in large structures like
tidal dwarf galaxies, but not in both.

HI observations indicate that tidal tails are gas rich
\citep{yun94,hib01a}, and many show evidence of in situ star formation
\citep{sch78,hib96,smi97,igl01,weil03}.
However, it is not clear what factors mediate tidal star formation.
Studies by \citet{ski86} and \citet{ski87} suggest that within the
disks of dwarf irregular galaxies it is the local HI surface density
rather than the fractional gas content of the galaxy that is of key
importance in producing high mass stars. Physically, the gas volume
density is more directly relevant to star formation than HI column
density. In disks the gas volume density and the HI column density
are closely related. However, in tidal debris geometric projection
effects may dilute that relationship.

In this paper, we examine whether there exists a relationship between 
super star clusters (SSCs) in tidal debris and the underlying HI
column density. Particularly, is the large scale (kpc-level) gas
structure responsible for SSC formation, or are the pc-scale cluster
formations unrelated to the kpc-scale gas structure? This is a step
towards understanding the variety of star formation modes in tails.

\section{Optical Data}

For this study, we use the {\it HST}/WFPC2 data on eight tidal
tail regions in four
advanced mergers observed in Cycle 7 (GO7466) and presented
by \citet{kni03}. The data were taken through the F555W and F814W
filters of {\it HST} (hereafter $V$ and $I$) and targeted the tails of the
Toomre Sequence mergers NGC 3256, NGC 3921, NGC 4038/39, and NGC 7252
\citep{tt77}. Observations were made between August 1998 and October 1999
for NGC 3256, NGC 3921, NGC 7252 and NGC 4038/39 (Pos B). Due to priority
scheduling for NICMOS, the Cycle 7 observations of two additional pointings
of NGC 4038/39 (Pos A and Pos C) were postponed to December 2000. Exposure
times for these observations were the same as that for Pos B (2000s and
1800s in $V$ and $I$ respectively centered on 12h 02m 0.6s, $-$18$\degr$ 55$\arcmin$ 32.3$\arcsec$ and
12h 02m 7.81s, $-$18$\degr$ 50$\arcmin$ 4.9$\arcsec$).

\begin{figure}
\plotone{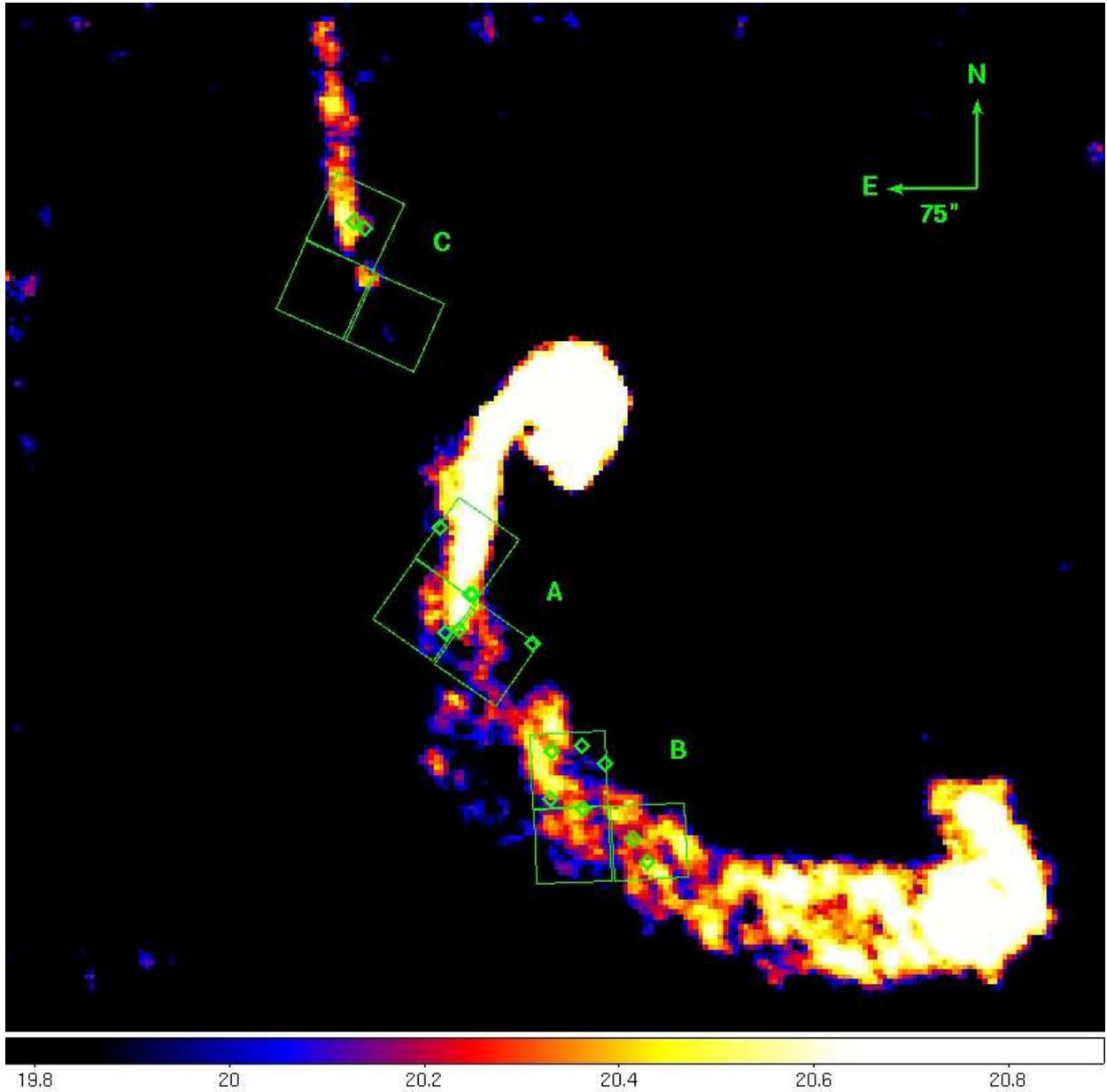}
\caption{Radio HI map of NGC 4038/9. 
The colours indicate the HI column density on a log scale, ranging from less than or equal to 19.8 (black) to greater than or equal to 20.6 (white).
The boxes show the locations of
the WFPC images of the system (PC not shown because it was not used)
and diamonds show the locations of the candidate SSCs with $M_V < -8.5$ and $V-I < 2.0$.}
\label{fig1}
\end{figure}

\begin{figure}
\plotone{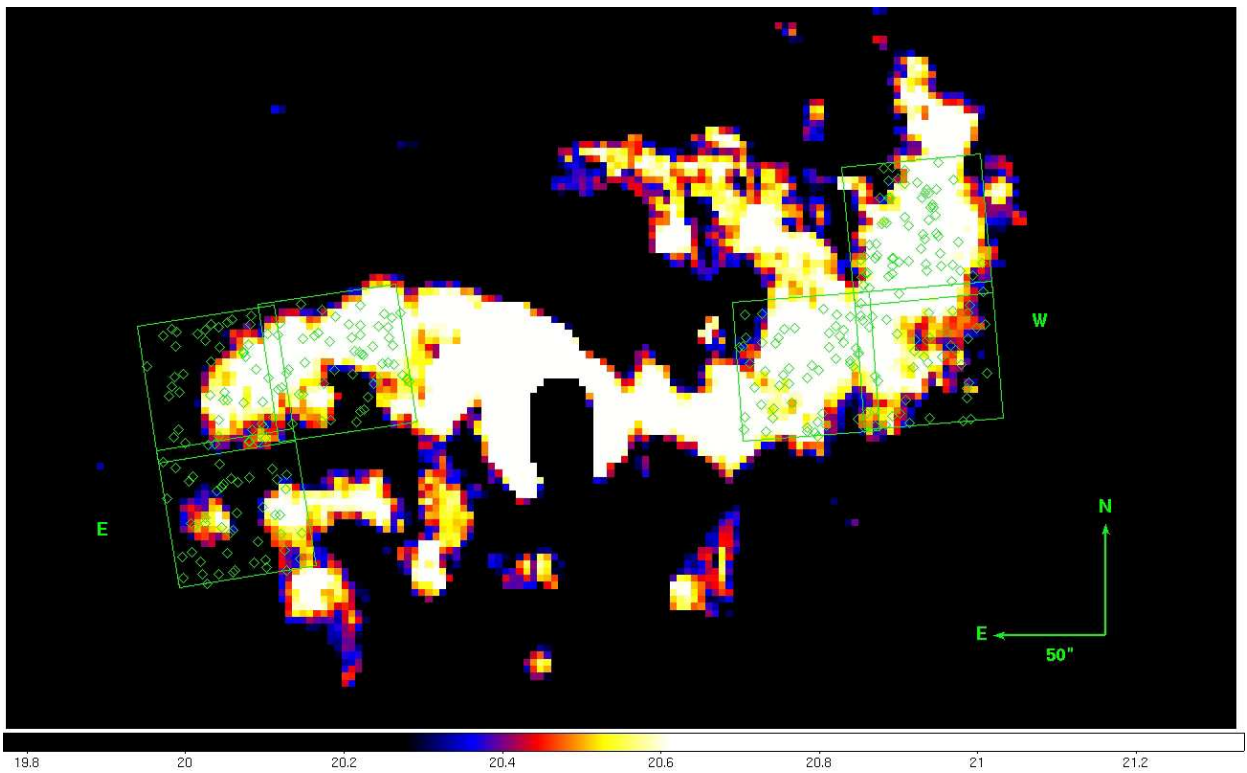}
\caption{Same as in Fig.1, but for NGC 3256.} 
\label{fig2}
\end{figure}

\begin{figure}
\plotone{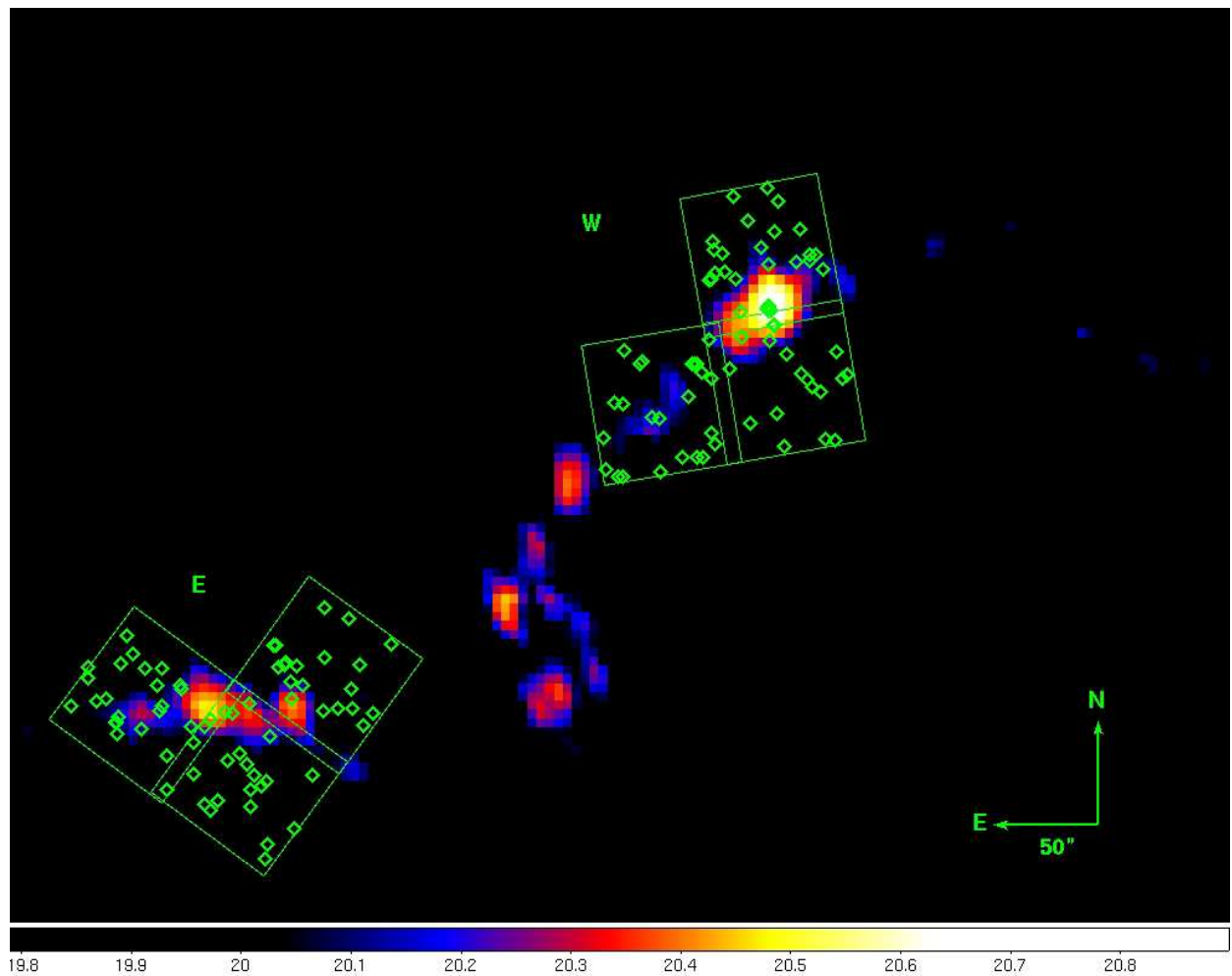}
\caption{Same as in Fig.1, but for NGC 7252.}
\label{fig3}
\end{figure}

\begin{figure}
\plotone{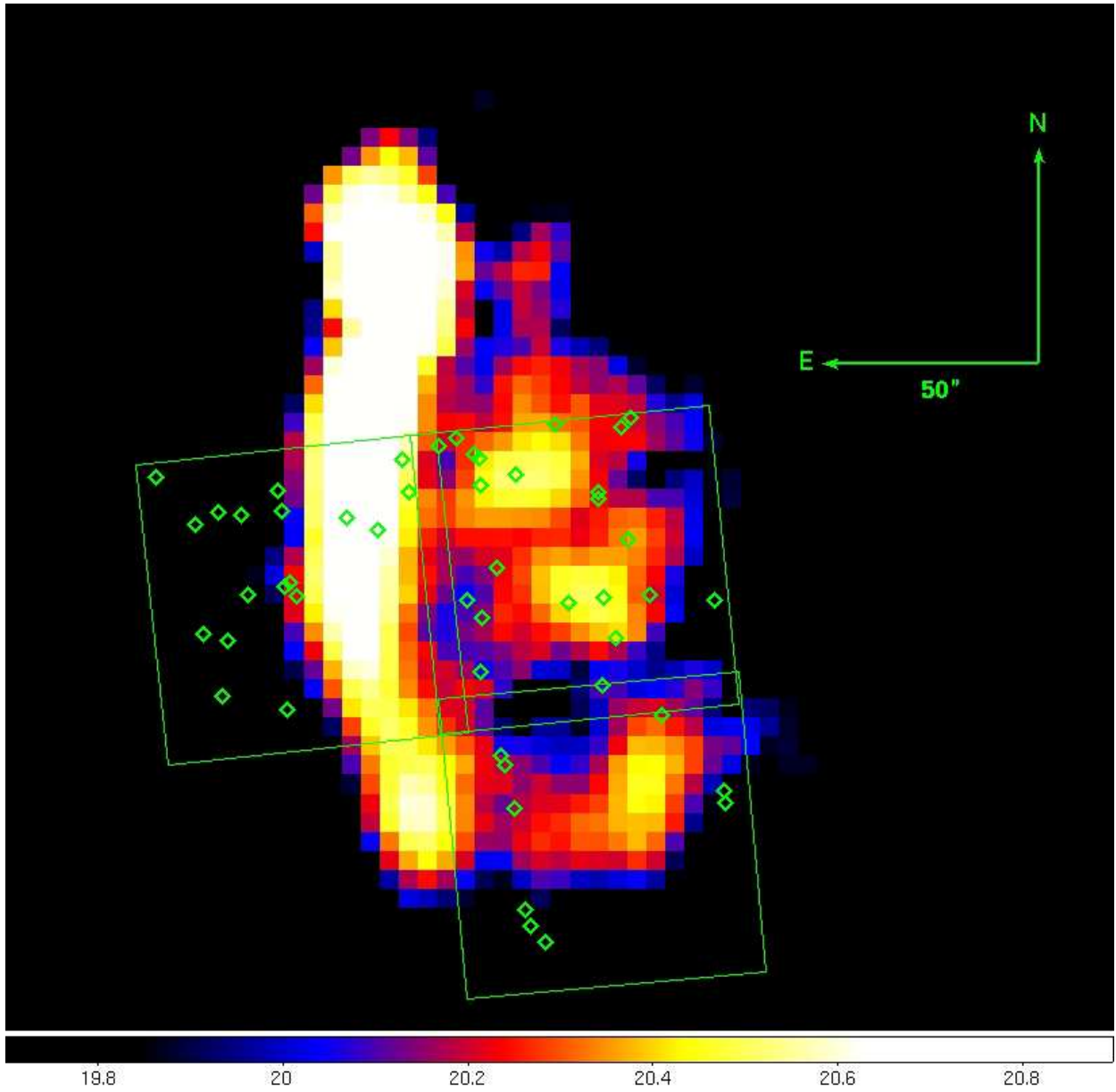}
\caption{Same as in Fig.1, but for NGC 3921.}
\label{fig4}
\end{figure}

Following the procedure described in \citet{kni03}, we automatically
detect point sources on pipeline-processed WFPC2 images in both $V$ and $I$
bands. Potential SSCs are identified by selecting only those point
sources with $V$ magnitude errors less than 0.25 mag. We only consider
sources with $M_V$ brighter than -8.5 (assuming them to be at the
distance of the tail) and bluer than $V-I < 2.0$ to eliminate
contamination by individual stars.
Figures~\ref{fig1}--\ref{fig4} show the positions observed in each tail
by WFPC2.
Limiting
magnitudes for each tail are given in Table 1. For the most distant
tail (NGC 3921), the completeness fraction is $\sim 50\%$ at the $M_V$ = $-$8.5
limit, determined by adding artificial stars to the WFPC2 image
\citep{kni03}. For slightly brighter sources, with $M_V$ = -9.0, our
sample in this most distant group is $>80\%$ complete.
Details of the total number of detected sources in-tail and out-of-tail\footnote {For each tail, the region in the $V$ image corresponding to
contiguous regions with $\sim 1$ count above the background was identified as
in-tail while all other regions were identified as out-of-tail.} and the number
of bright sources selected  are given in Table 2. These were computed assuming
an area of 5.3 sq. arcmin for the three WFPC2 chips
and using the distance given in Table 1.

\section{Radio Data}

The 21 cm HI maps were obtained from various sources.
The radio maps for NGC 4038/9, NGC 7252, and NGC 3921 were
obtained using the VLA in the C+D array configuration, with a
resolution $\sim$20\arcsec\ (see \citet{hib01b,hib94} and \citet{hib96} respectively).
HI maps for NGC 3256 were obtained using
the ATCA in 3 different antenna configurations by \citet{eng03}.  The
FWHM of the beam, its corresponding physical size (ranging from 1.4
kpc to 8.3 kpc for different pairs), and the limiting HI column density
for the four galaxies are given in Table l.
All HI column densities are measured from zeroth moment maps made from
the AIPS tasks MOMNT \citep{rupen99}. 

\section{Relationship between SSCs and HI}

Figures~\ref{fig1}--\ref{fig4} present images of the
integrated HI line emission for the four mergers, showing the outline
of the WFPC2 pointings and location of detected SSC candidates.
The SSCs are parsec scale in size while the HI data has a few kpc
scale resolution. Though it is not possible to study the gas and the
clusters on the same scale, it is nevertheless interesting to explore
whether the cluster environment is rich in HI. In order to study the
correlation between the point sources derived from the photometry of
the WFPC images and the distribution of HI, we first converted the
positions of the point sources (both in-tail and out-of-tail) from
pixel coordinates on the chip to right ascension and declination.
The HI maps are then used to obtain the HI column density at these
coordinates.

\begin{figure}
\plotone{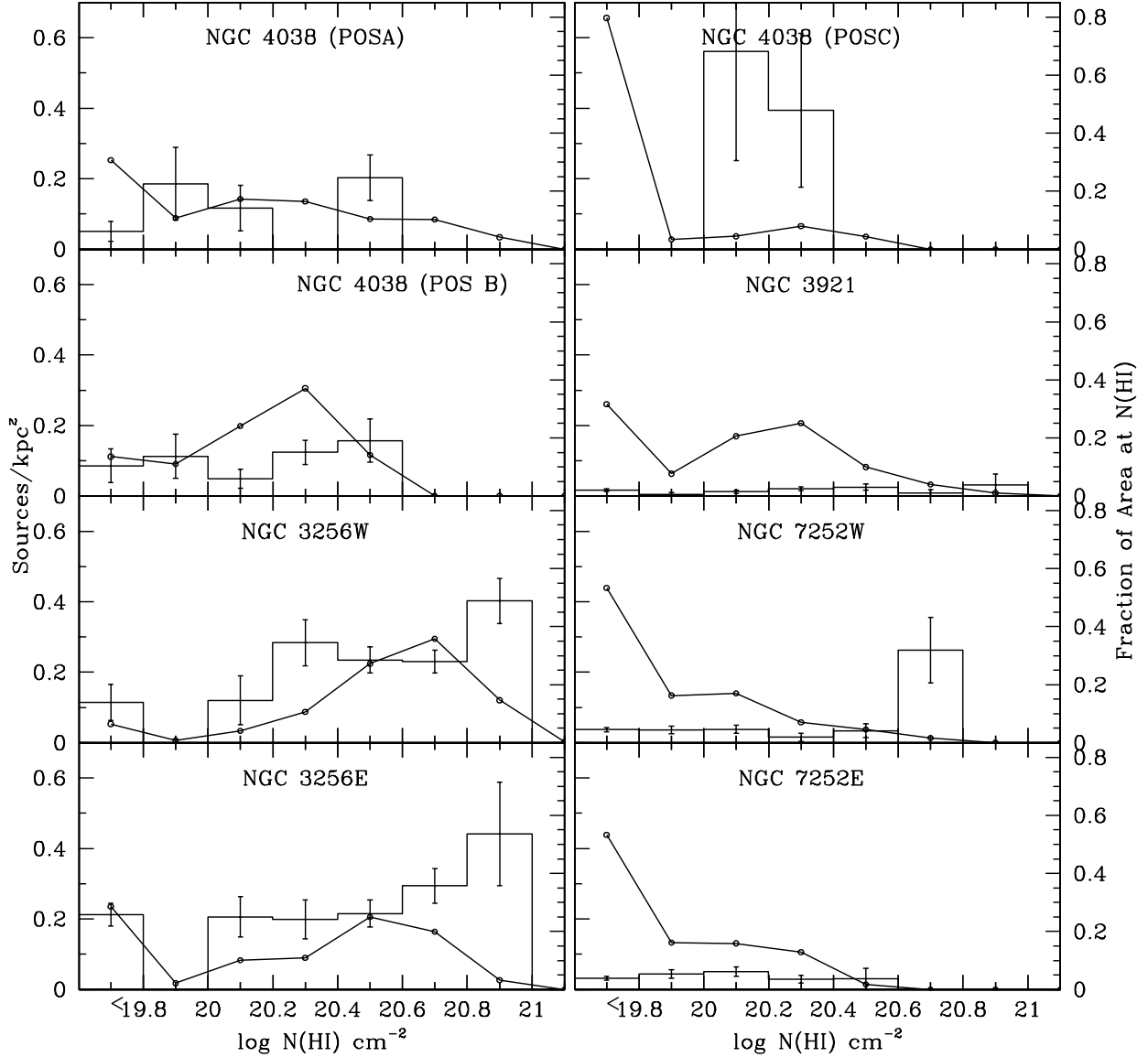}
\caption{Histograms of source density for the tail regions in the four
pairs (on the left-hand scale) as a function of the neutral hydrogen
column density. The curves show the fraction of the WFPC2 field-of-view
(on the right-hand scale) within each column density bin. The
neutral hydrogen column density on the horizontal scale is given in
logarithmic units.  The fractional area curves indicate the expected
distribution for randomly distributed foreground/background sources.}
\label{fig:fig5}
\end{figure}

Figure~\ref{fig:fig5} plots histograms of the surface density of star
clusters (number per square kpc) as a function of the underlying HI
column density. The first bin on each graph shows the clusters
detected in regions with no observed HI flux (down to the limiting
values listed in Table 1), effectively setting a background level for
point sources unrelated to the tidal tail.
Also plotted in each
panel is the fractional area subtended by WFPC2 pixels in each column
density bin (solid curve). This curve represents the expected
detection probability for a uniform distribution of
background/foreground sources.  If the observed sources follows this
curve it implies that there is no significant population of SSCs associated
with the tail.

From Figure~\ref{fig:fig5} we draw the following conclusions:
(1) three of the eight tails --- NGC\,3256W, NGC\,3256E and NGC\,7252W
--- have a significant population of SSCs, but only at high HI column
densities ($\log N_{HI} \gtsim 20.6$ cm$^{-2}$); (2) of the five remaining tails,
three --- NGC\,4038/39 Pos B, NGC\,4038/39 Pos C, and NGC 7252E --- have
few, if any, SSC candidates and little or no gas at high column density
($\log N_{HI} \gtsim 20.6$ cm$^{-2}$); (3) the two remaining tails ---
NGC\,4038/39 Pos A, NGC\,3921 --- have gas at high column density, but
no significant population of SSCs.

\begin{figure}
\plotone{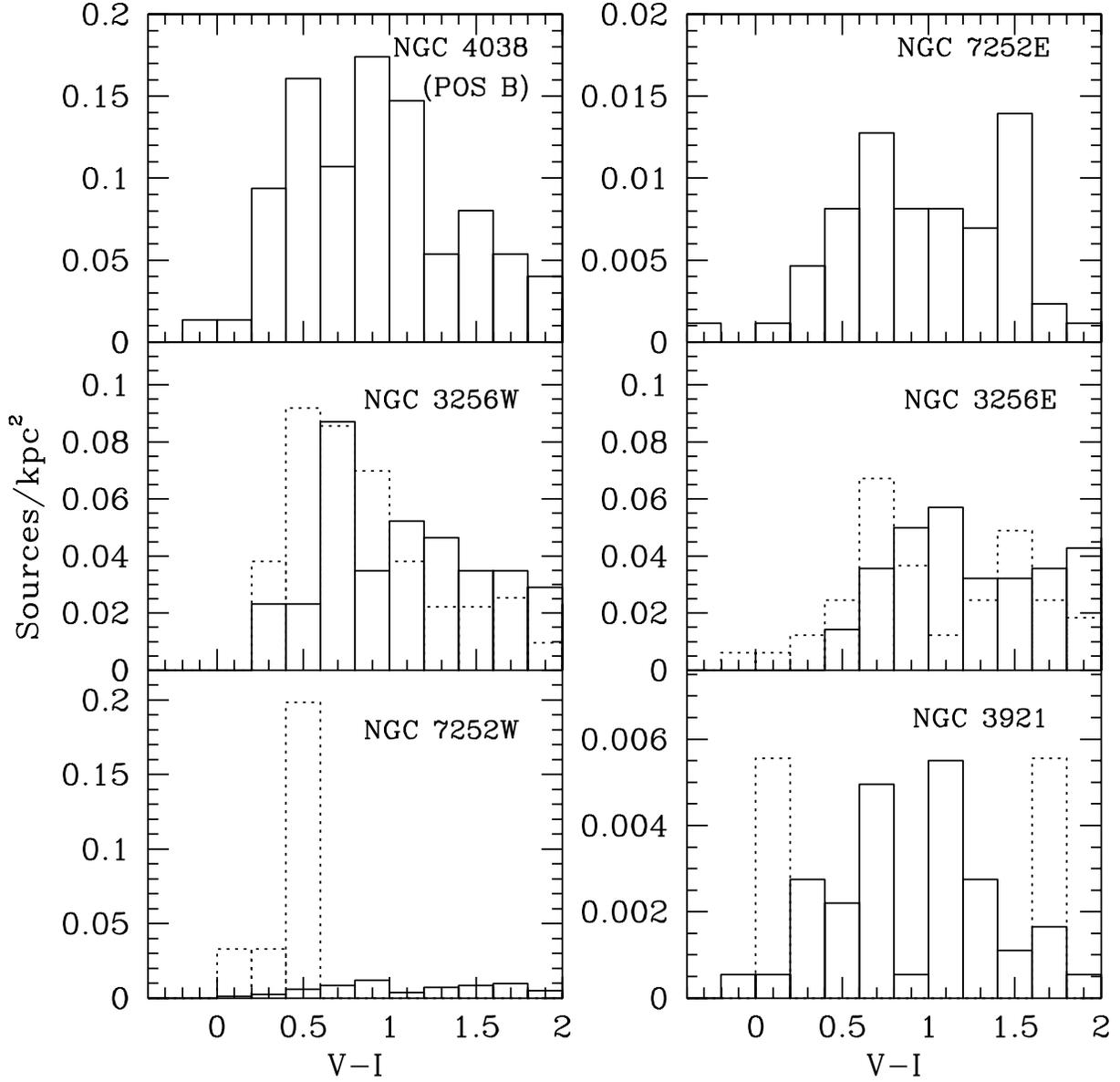}
\caption{Number of sources per unit area ($kpc^{-2}$) with colour $V-I$ for sources detected, separated by whether they are in regions with $\log N_{HI} > 20.6$cm$^{-2}$ (dotted line) or with $\log N_{HI} < 20.6$cm$^{-2}$ (solid line). Areas are calculated separately for regions with $\log N_{HI} > 20.6$cm$^{-2}$ and $\log N_{HI} < 20.6$cm$^{-2}$ so that the number of sources in the bin are normalized by the area above or below the column density threshold.
NGC 4038/39 Pos A and Pos C are
not shown because no sources were detected.  Dotted lines are not shown for
NGC 4038 Pos B and for NGC 7252E because these tails had no regions with
$\log N_{HI} > 20.6$cm$^{-2}$.  A bluer colour for the dotted histograms indicates
a real population of young star clusters in high HI column density regions.
}
\label{fig:fig6}
\end{figure}

If, as we propose, the excess of SSCs in regions with $\log N_{HI} >
20.6$ cm$^{-2}$ reflects a true excess, we would expect the colour distribution
of these objects to differ from those of sources outside of the tail.
For this comparison we eliminate sources redder than $V-I=2.0$, since
these are very unlikely to be real sources.  Figure~\ref{fig:fig6}
shows histograms of the distributions of ($V-I$) colour,
displayed separately for sources in regions with $\log N_{HI} > 20.6$ cm$^{-2}$
and for sources in regions with $\log N_{HI} < 20.6$ cm$^{-2}$. Source
densities in each bin are presented per unit area with that column density
so that the distributions can be easily compared.
For each tail,
we perform a Kolmogorov-Smirnov (K-S) test \citep{pre86} to evaluate
whether sources within regions of high HI column density ($\log N_{HI}
> 20.6$ cm$^{-2}$) are drawn from the same parent population as sources drawn
from low HI column density ($\log N_{HI} < 20.6$ cm$^{-2}$).  The colours of
sources in high column density regions of NGC 3256W differ significantly
from those
in lower column density regions, with a probability of only $0.12$\%
that they are drawn from the same distribution.  The sense of the
difference, that the sources in high column density regions are bluer,
is what we would expect if they are real SSCs.  We find a similar
result for NGC 7252W, with only a $0.03$\% chance that the colours of
the sources in the high and low $N_{HI}$ sources are drawn from the
same distribution.  NGC 3256E did not show a significant difference
between the $V-I$ distributions of the two populations of sources,
with a probability of 15\% that they are drawn from the same
distribution.  This indicates that in that case there may be a
significant contamination from background/foreground sources that
happen to fall in an area of high HI column density. NGC 3921 had only
2 sources in the high HI column density regions, so in that case we could
not draw any conclusion about differences in the colour distributions of
high and low HI column density sources.

An estimate of the density of SSCs in areas with $\log N_{HI} > 20.6$ cm$^{-2}$
is made by taking the observed number per unit area in these
regions and subtracting the expected surface density of background/foreground sources.
The expected background/foreground for each region is tabulated in Table 3,
estimated in two
ways: 1) by dividing the total number of sources in the entire field
by the area of the field; 2) from the first bin of each panel of
Fig.~\ref{fig:fig5}, i.e.  taking the density of sources in regions
with $\log N_{HI} < 19.8$ cm$^{-2}$, since these sources are highly unlikely to
be related to the tails. The first method is an overestimate since it
includes any real SSCs that may lie in the field.  The second method
is not biased, but suffers from errors due to small number statistics
in fields without much area with low $N_{HI}$.  In Table 3, we list
the number of candidate SSCs and the subtended area in kpc$^2$ for
regions with $\log N_{HI}
> 20.6$ cm$^{-2}$ in Cols.(2) and (3), the observed source density in Col.(4)
obtained by dividing Col.(2) by
Col.(3), the estimated background source density in Col.(5) obtained using the first method given above,
the estimated source density of candidate SSCs above the background in Col.(6) obtained using the background from Col.(5),
the estimated background source density in Col.(7) obtained using the second method given above,
the estimated source density of candidate SSCs above the background in Col.(8) obtained the background from Col.(7).
  and our estimates of the resulting background subtracted SSC
densities in the $\log N_{HI} > 20.6$ cm$^{-2}$.  Uncertainties are calculated
assuming Poisson statistics for the number of detected sources.

If $\log N_{HI} > 20.6$ cm$^{-2}$ was the only condition necessary for SSC
formation we would expect the SSC densities in all such tail regions
to be the same.  We do find that the values for the SSC density in
high $N_{HI}$ regions are consistent for NGC 3256W, NGC\,4038/39 Pos A,
NGC 7252W, and NGC 7252E. The fact that the same N(HI) threshold applies
for these four regions suggests that projection effects are not
significant or are the same for these four tail regions. That is, the
gas volume density is likely to be similar for them so that the number
of clusters per unit area (also a projected quantity) is similar.
Although no SSC sources were observed for
NGC\,4038/39 Pos A, it is also true that there is only 11.2~kpc$^2$ of
area in that tail region with $\log N_{HI} > 20.6$ cm$^{-2}$.  Even if it had an
SSC density as high as that of NGC 3256W (by method 2) we would expect
only $3.1\pm0.7$ SSCs, so detecting none is perhaps not too surprising ($5\%$
probability).

However, NGC 3921 is an exception to the rule of having enhanced SSC
formation in regions of high $N_{HI}$.  After background subtraction,
it has no SSCs detected in its $\log N_{HI} > 20.6$ cm$^{-2}$ regions, with an
SSC density in these regions differing from that of NGC3256W at
a $4.5\sigma$ level.  Based upon its area at high $N_{HI}$, 180~kpc$^2$,
and the SSC density of NGC 3256W, we would expect $50\pm11$ SSCs in
the NGC 3921 field, and we see only $2$.  Furthermore, the $2$ detected sources
are consistent with the expected background level, such that we estimate
that no SSCs are detected.  Using Poisson statistics, the probability
of measuring 0 when 50 are expected is $2\times 10^{-22}$.  
We conclude that, despite having a large
area that exceeds the $N_{HI}$ threshold, NGC 3921 appears to be
unable to form clusters. This suggests that some other
factors influence the formation of star clusters, and that having
significant areas with $N_{HI}$
above the threshold value is a necessary but not sufficient
condition. One possible explanation is that the projected HI column
density in NGC 3921 may exceed the threshold because of a large line of sight
extent, such that the physical gas volume density is actually low. However,
the optical source density would be subject to the same projection effect.
From the location of the WFPC2 pointing and the tidal tail morphology
and kinematics \citep{hib96}, this region of the tail should lie mostly
in the plane of the sky, where projection strong effects are not
expected \citep{bournaud2004}.

\section{Discussion \& Conclusions}

Formation of stars in tidal debris, whether in the form of isolated
clusters or tidal dwarfs, preferentially occurs in regions with larger
HI column densities (on kpc scales). The HI column density is the best
indicator we have of the more physically relevant quantity, the gas
volume density. In this study, we find a threshold
value of $\log N_{HI}\gtsim 20.6$ cm$^{-2}$ (on kpc scales) for regions of
increased SSC formation.

Our study of four pairs of interacting galaxies shows populations
of SSCs that have formed in regions with $\log N_{HI} \gtsim 20.6$
cm$^{-2}$ in NGC 3256W, NGC 3256E, and NGC 7252W. Similarly,
there appear to be few clusters formed in regions with lower $\log
N_{HI}$ values. In NGC\,4038/39 Pos A, Pos B, and NGC 7252E there are no
substantial populations of SSCs, but this is consistent with
their relatively small areas with $\log N_{HI}$ $\gtsim$20.6cm$^{-2}$.
For NGC 3921, however, there is a significant area exceeding
this $N_{HI}$ threshold, and no SSCs are found.

Since NGC 3921 is the most distant group, we should consider biases
that might have affected this conclusion.  For example, our 50\%
detection limit of $M_V$ = $-$8.7 for NGC 3921 is not as faint as for
the other tails (see Table 1).  However, since we have only considered
point sources brighter than $M_V$ = $-$8.5 in our analysis, we could
not possibly be missing more than half the sources in NGC 3921. Thus
the fact that we do not detect any SSCs compared to the exception of
$50\pm11$ is still highly significant.  Another possible bias arises
because the beamsize is largest for this most distant tail, which
would lead to more dilution of small, high column clumps of HI so that
some regions that exceed the $N_{HI}$ threshold might be missed.
However, our conclusion is that there {\it are} significant high HI
column density areas observed in NGC 3921, but that they do not have
SSCs.  If there were even more high HI column density areas with no
detected SSCs that would only strengthen this conclusion.

Based on a stability analysis for a self-gravitating gas disk,
\citet{ken89} showed that there should be a threshold $N_{HI}$
value below which star formation is strongly suppressed.
\citet{ski86} and \citet{ski87} observationally determined
a threshold of $\log N_{HI} = 21.0$ cm$^{-2}$, averaged over 500~pc within the
disks of dwarf irregular galaxies.  They also found that giant
extragalactic HII regions are only located in regions where the HI
surface densities are a factor of 3 to 5 times this threshold.
Recently, \citet{wal06} reported the discovery of HII regions in the
tidal arms of NGC 3077 found only where the HI column density reaches
values above $\log N_{HI} = 21.0$  cm$^{-2}$, averaged over 200~pc. Observations in
the far UV by the GALEX satellite find a close correspondence between
HI column density and FUV emission when the HI column density reaches
a threshold of 2.5$\times 10^{20} {\rm cm}^{-2}$
\citep{neff05,thilker05,gil05}.  \citet{mun04} find a HI cloud at the
base of the northern tail in the interacting system NGC 3227/3226,
associated with massive ongoing star formation.  This is seen as a cluster of
blue knots (M$_B$ $<\approx$ 15.5), cospatial with a ridge of high
neutral hydrogen column density ($\log N_{HI}$ $\approx$ 3.7 $\times$
10$^{21}$ cm$^{-2}$). A similar result is reported by \citet{hor04} in
the tidal regions of the interacting galaxy NGC 6872.  These results
are all generally consistent with the value of $\log N_{HI} \approx 20.6$
cm$^{-2}$ that we have found for SSCs in the tails of interacting galaxies.  Small
differences can be due to differences in the scales over which the
$N_{HI}$ is determined.

The measurement of $\log N_{HI}$ for this study is obtained by
averaging over a large beam size. In principle, correlations between
point sources and $N_{HI}$ might be significantly stronger and the
threshold much larger if the HI column density could be examined on
scales more similar to the size of a star cluster.  The amount and the
scale on which HI is present and its distribution in these regions of
high HI column densities could influence the scales of star
formation. Thus, we cannot yet compare thresholds for formation of
tidal dwarfs, SSCs, loose stellar associations, and HII regions.

The excess of point sources found in the western tail of
NGC 7252 is of particular interest.  The high detection rate found at
$\log N_{HI} > 20.6$ cm$^{-2}$ in this case is due entirely to star clusters
associated with its tidal dwarf galaxy (TDG) candidate
\citep{hib94}. A similar result was found by \citet{sav04} for the TDG
candidate in NGC 4038, with the three brightest stellar associations
in its vicinity ($M_V < -8.0$) in regions with $\log N_{HI}$ $>$ 20.6 cm$^{-2}$.

If a threshold value of HI column density was the only factor
determining the source density of star clusters, the number of real
star clusters detected after the subtraction of background sources
would be consistent with the average source density for high HI
column density regions.
We find that our estimates of the number of real sources do
agree quite well for most tails in our sample.  However, there is a
glaring discrepancy in the case of NGC 3921 in which we would expect
to find about 50 clusters but none are detected (after background/foreground
subtraction). This could be the result of a large projection effect
in the case which pushed N(HI) above the threshold despite a relatively
low gas volume density. In addition to the star
formation threshold, other factors like pressure
\citep{jog92,jog96,elm97} or the strength of the encounter
\citep{kee03} could also influence the type of stellar associations that
are formed. \citet{elm97} show that differences in pressure at the
time of formation gives rise to the primary structural differences
between various cluster types. They find that large-scale shocks, as
seen in interacting galaxies, could lead to regions of high pressure
which are conducive to the formation of globular clusters.  Perhaps,
the NGC 3921 debris did not experience these high pressure conditions.

Future studies with observations of a larger sample of mergers,
encompassing the entire tidal debris regions, combined with higher
resolution HI observations, are needed. Such a sample will facilitate
comparison of HI thresholds for different types of cluster and dwarf
formation. It would also allow us to isolate other factors that
influence formation of various kinds of stellar associations.

\acknowledgments
We would like to thank the referee, Peter Weilbacher for
helpful comments. This work was supported at Penn State by grant \#0071223 from the National
Science Foundation.  JM also received support from an REU
Supplement to that grant.

{}

\clearpage

\begin{table}
\scriptsize
\begin{center}
\caption{Comparison of radio and optical data}
\label{tb1}
\begin{tabular}{lccccc}
\tableline\tableline
Pair & Distance & \multicolumn{2}{c}{Radio Beam} & $\log N_{HI,min}\tablenotemark{a}$ & M$_{V,50\%}\tablenotemark{b}$\\
& (Mpc) & FWHM & Size (kpc) & (cm$^{-2}$) & (mag) \\
\tableline
NGC4038/39 &13.8&20.7\arcsec$\times$15.4\arcsec&1.4$\times$1.0&19.7& -5.8\\
NGC3256 &37.6&25.7\arcsec$\times$19.3\arcsec&4.6$\times$3.5&20.0&-7.5\\
NGC7252 &64.4&26.9\arcsec$\times$16.1\arcsec&8.3$\times$5.0&19.7&-8.1\\
NGC3921 &80.3&19.3\arcsec$\times$18.1\arcsec&7.5$\times$7.1&19.7&-8.7\\
\tableline
\tablecomments{Physical scales computed adopting H$_o$= 71 km s$^{-1}$ Mpc$^{-1}$, except for NGC4038/39 where we adopt a distance of 13.8 Mpc from \cite{sav04}}
\tablenotetext{a}{minimum detectable HI column density}
\tablenotetext{b}{50\% completeness limit from \cite{kni03}.}
\end{tabular}
\end{center}
\end{table}

\begin{deluxetable}{lccccccc}
\tabletypesize{\scriptsize}
\tablecaption{Source Statistics}
\tablehead{\colhead{Tail} & \colhead{Total} & \colhead{Bright\tablenotemark{a}} &
\colhead{Bright \& Blue\tablenotemark{b}} & \colhead{Total} & \colhead{Bright\tablenotemark{a}} & \colhead{Bright \& Blue\tablenotemark{b}} & \colhead{Fractional area} \\
\colhead{} & \colhead{In Tail} & \colhead{In Tail} & \colhead{In Tail} & \colhead{Out of Tail} & \colhead{Out of Tail} & \colhead{Out of Tail} & \colhead{with $\log N_{HI} < 19.8$ cm$^{-2}$}}
\startdata
NGC3256E & 105 & 74 & 47 & 95 & 73 & 51 & 0.28\\
NGC3256W & 127 & 71 & 57 & 129 & 86 & 56 & 0.06\\
NGC7252E & 29 & 27 & 23 & 44 & 37 & 33 & 0.53\\
NGC7252W & 37 & 35 & 31 & 42 & 38 & 30 & 0.53\\
NGC3921 & 16 & 16 & 14 & 33 & 33 & 30 &0.31\\
NGC4038B & 47 & 6 & 4 & 27 & 3 & 3 & 0.13\\
NGC4038A & 145 &18 & 3   &253  &51 & 3 & 0.30   \\
NGC4038C & 16 &1  &1 & 124 & 1 &1 & 0.79   \\
\enddata
\tablenotetext{a}{$M_V < -8.5$}
\tablenotetext{b}{$M_V < -8.5$ and $V-I < 2.0$}

\label{tb2}
\end{deluxetable}

\begin{table}
\scriptsize
\begin{center}
\caption{Comparisons of real source densities for different tails}
\label{tb3}
\begin{tabular}{lccrrrrrr}
\tableline\tableline
(1)&(2)&(3)&(4)&(5)&(6)&(7)&(8)\\
Tail & Obs. Src.& Area $>$ 20.6 & Obs. Src. & Backgr. 1 & Est. SSC1 & Backgr. 2 &Est. SSC2 \\
     &    \#       & [kpc$^2$]  &  [kpc$^{-2}$]    & [kpc$^{-2}]$ & [kpc$^{-2}]$ & [kpc$^{-2}$] & [kpc$^{-2}$] \\
\tableline
NGC 3256E& 65  & 163.6 & 0.40$\pm$0.05&0.23$\pm$0.02&0.17$\pm$0.05&0.21$\pm$0.03&0.18$\pm$0.06\\
NGC 3256W& 161 & 409.5 & 0.39$\pm$0.03&0.25$\pm$0.02&0.15$\pm$0.04&0.11$\pm$0.05&0.28$\pm$0.06\\
NGC 7252E& 0 & 0.0 & 0.0&0.04$\pm$0.01&-0.04$\pm$0.01&0.03$\pm$0.01&-0.03$\pm$0.01\\
NGC 7252W& 8 & 30.2 & 0.26$\pm$0.09&0.04$\pm$0.01&0.22$\pm$0.09&0.04$\pm$0.01&0.23$\pm$0.09\\
NGC 3921& 2 & 179.9 & 0.01$\pm$0.01&0.017$\pm$0.003&-0.005$\pm$0.008&0.017$\pm$0.004&-0.006$\pm$0.01\\
NGC 4038B& 0 & 0.0 & 0.0&0.06$\pm$0.03&-0.06$\pm$0.03&0.05$\pm$0.05&-0.05$\pm$0.05\\
NGC 4038A&0 & 11.2 & 0.0&0.07$\pm$0.03&-0.07$\pm$0.03&0.028$\pm$0.028&-0.028$\pm$0.028\\
NGC 4038C&0 & 0.0 & 0.0&0.02$\pm$0.01&-0.02$\pm$0.01&0.0&0.0\\
\tableline
\tablecomments{(1): Identification, (2): Number of candidate SSCs with $M_V < -8.5$ and $V-I < 2.0$ in regions with column density $\log N_{HI} > 20.6$cm$^{-2}$, (3): Area on each image for which $\log N_{HI} > 20.6$cm$^{-2}$, (4): (2) divided by (3) which gives the observed source density, (5): Background \#1 estimated by dividing the total number of sources in the entire field by the area of the field, (6): Estimated number of candidate SSCs above the background, if value of background is taken from (5), (7): Background \#2 estimated by taking the density of sources in regions with $\log N_{HI} < 19.8$cm$^{-2}$. This is the value of the first histogram bin from each panel in Fig.~\ref{fig:fig5}, (8): Estimated number of candidate SSCs above the background, if value of background is taken from (7)}

\end{tabular}
\end{center}
\end{table}

\clearpage

\clearpage
\end{document}